# Etude et développement d'un protocole dymétrique pour sécuriser les communications des RCSF

Yassine Maleh et Pr. Abdellah Ezzati
Laboratoire Veille et Technologies Emergente (LAVETE)
Faculté des Sciences et Techniques de Settat, Maroc
y.maleh@uhp.ac.ma, abdezzati@uhp.ac.ma

**Résumé :**
Durant cette dernière décennie, les réseaux de capteurs sans fil (RCSF) ont attiré l'attention des chercheurs et des services de recherche et développement en raison de leur facilité de déploiement et de leur champs d'application dans divers domaines, y compris la sécurité et la surveillance, le contrôle, la maintenance des systèmes complexes, l'agriculture, e-santé, etc. Toutefois, en raison des ressources limitées de capteurs et les environnements de déploiement hostiles, ce type de réseaux est vulnérable à plusieurs types d'attaques. La sécurité est l'un des défis les plus cruciaux que les RCSF peuvent confronter. Les réseaux de capteurs sont très vulnérables contre les attaques. Par conséquent, il est nécessaire d'utiliser des mécanismes efficaces pour protéger ce type de réseau. Les mécanismes de cryptographie assurent la sécurité et ses services et jouent un rôle très important dans la détection et la prévention contre les attaques. Dans ce travail, nous nous concentrons dans un premier temps sur l'étude et l'analyse des algorithmes de cryptographie pour les RCSF. Par la suite, nous proposons un nouveau prototype léger à base de LEAP+ pour sécuriser les communications. Notre contribution est destinée aux applications utilisant les nœuds avec une faible mobilité et nécessitant un niveau de sécurité élevé. Contrairement à la plupart des méthodes proposées dans la littérature pour des topologies spécifiques, notre prototype de sécurité peut couvrir à la fois les besoins des topologies plats et hiérarchiques. L'évaluation de notre solution a été effectuée en utilisant le simulateur de TOSSIM. Les résultats ont montré que notre schéma assure le passage à l'échelle, résiste contre la majorité des attaques, avec une consommation d'énergie faible. Enfin, la complexité du protocole est analysée et comparer avec d'autres systèmes symétriques.

**Mots-clés :** réseau de capteurs sans fil, la cryptographie, gestion des clés, l'établissement de clés, LEAP+, TOSSIM

## I. Introduction

Un réseau de capteurs sans fil (RCSF) est un ensemble de nœuds dont la fonction principale est la collecte d'information, capables de communiquer entre eux dans le but de réaliser diverse tâches. La facilité de déploiement de ce type de réseau constitue un atout qui les rend facilement intégrables dans une grande variété d'applications comme les applications militaires, environnementales, domotiques, industrielles, etc. En raison de leur petite taille, cela permet leur déploiement dans des endroits à accès difficile comme les champs de bataille [1]. Les RCSF représentent une nouvelle perspective pour un grand nombre d'applications qui concernent l'environnement. Ils sont essentiellement utilisés pour la récolte des données comme par exemple celles de phénomènes physiques, de la surveillance des évolutions de terrains, etc. La figure 1 montre un exemple d'application des RCSF dans le domaine militaire.

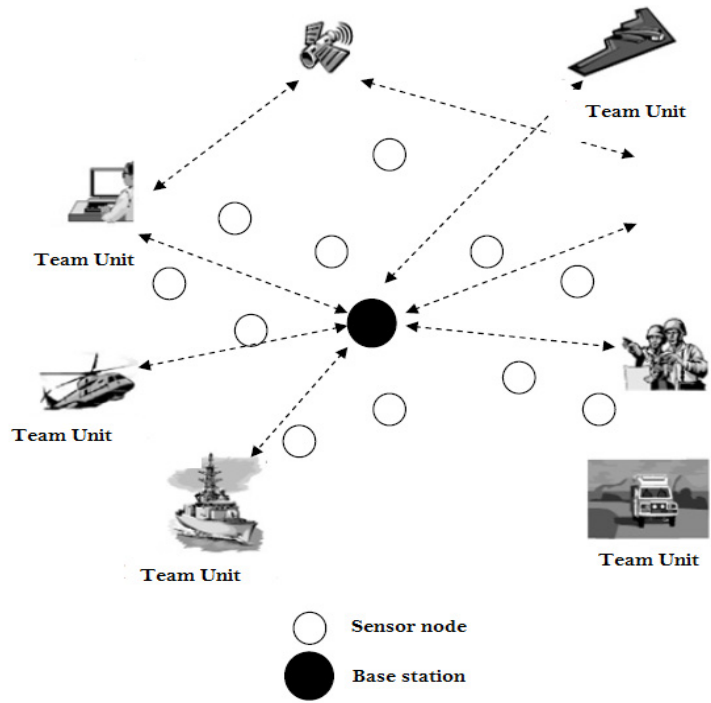

Figure 1. Application des RCSF dans le domaine militaire

Les différents caractéristiques des réseaux de capteurs à savoir la limitation en capacité de calcul et de mémoire ainsi qu'en ressources énergétiques. Ces contraintes représentent un véritable challenge pour les chercheurs quand il s'agit de concevoir des solutions répondant aux besoins que nous venons d'évoquer. Un des objectifs les plus essentiels est l'économie de l'énergie des nœuds qui a pour but de prolonger la durée de vie d'un tel réseau. Les protocoles nécessaires au fonctionnement des RCSF comme les protocoles d'accès au médium, de routage et de sécurité doivent prendre en compte les contraintes des nœuds du réseau tout en économisant le plus possible leur consommation d'énergie. Le but des nouvelles propositions et des nouveaux protocoles dédiés aux RCSF est de faire un compromis entre la qualité du service rendu par ces solutions et le respect des contraintes imposées par les nœuds.

Par conséquent, il est nécessaire d'utiliser des mécanismes efficaces pour protéger ce type de réseau. Toutefois il est bien connu, que les systèmes de cryptage sont des mécanismes de sécurité très efficaces pour protéger le réseau contre les attaques de tout type. Par ailleurs, les techniques de cryptographie doivent être conçues pour détecter l'exécution des attaques les plus dangereuses. En outre, ces techniques doivent être légères pour s'adapter à la nature des ressources limitées du RCSF.

Les nœuds de capteurs sont limités en termes de calcul, de mémoire et des capacités énergétiques, ces contraintes influencent négativement le bon fonctionnement des techniques intelligentes spéciales qui fournissent la sécurité requise. Les protocoles de gestion de clés assurent des chemins sûrs dans les réseaux de capteurs. L'idée est d'avant d'ouvrir la voie en toute sécurité, les nœuds doivent partager certaines informations d'identification et d'authentification. Nous appelons ces informations d'identification de sécurité des clés. Ces clés de sécurité ont besoin eux aussi d'un système de gestion qui prend la responsabilité de la création, le maintien et la distribution de clés de sécurité [2].

Dans ce travail nous proposons un prototype de sécurité globale pour les RCSF basé sur le fameux protocole LEAP+, proposé par Zhou. Ce protocole est conçu pour sécuriser les communications réseau avec comme premier objectif premier est de limiter immédiatement l'impact de la sécurité d'un nœud compromise sur le réseau. Le protocole supporte quatre types de clés qui sont échangés afin de répondre aux exigences différentes de la sécurité des RCSF :

- Clé individuelle (Individual Key) : Partagée avec la station de base
- Clé par-paire (Parwise Key) : Partagée avec un autre nœud de capteur
- Clé de cluster (Cluster Key) : Partagée avec plusieurs nœuds voisins
- Clé Global (Global Key) : Partagée par tous les nœuds du réseau.

L'hypothèse sur laquelle LEAP+ se repose est qu'il existe une limite inferieure sur un intervalle de temps Tmin nécessaire pour qu'un adversaire puisse compromettre un nœud. Cette hypothèse parait pratique, mais seulement dans des conditions idéales, il se peut que Tmin soit plus grande que celle assumée. Pour résoudre cette limite, nous proposons deux modèle, le premier repose sur l'utilisation d'une vérification périodique "Periodic Chek" pour détecter le nœud compromise, La deuxième méthode est d'exécuter un numéro de séquence dans chaque nœuds et les comparés après l'établissement de clé par-paire avec les informations stockés dans la station de base BS, qui prend ensuite la décision sur la suppression ou non des clés partagés. Nos évaluations ont été implémentées sous le simulateur TOSSIM, et donnent une idée précise et détaillée du surcoût de la consommation de ressources nécessaire pour assurer le niveau élevé de sécurité attendu.

Ce travail sera organisé comme ainsi. Dans la deuxième partie, nous proposons les objectifs de la sécurité pour les RCSF, ensuite nous identifions les différents attaques et vulnérabilités dans les RCSF, en soulignant les remèdes employés pour résister à chacune d'entre-elles. Nous présentons ensuite un état de l'art en termes d'algorithmes de gestion de clés pour les RCSF. Par la suite, nous discuterons notre modèle de sécurité pour les RCSF. Nous évaluons notre proposition en termes de temps de génération de clé, de connectivité, et de résistance aux attaques et nous la comparons à d'autres méthodes proposées dans la littérature. Ce topic se termine par une conclusion générale et un ensemble de perspectives.

## II. Les objectifs de la sécurité pour les RCSF

Un réseau de capteurs est un type particulier de réseau. Il partage certaines similitudes avec un réseau informatique typique, mais pose également des exigences uniques. Nous pouvons classer les objectifs de la sécurité d'un RCSF en deux objectifs : principal et secondaire. Les principaux objectifs comprennent des objectifs de sécurité qui devraient être disponibles dans tout système (La confidentialité, la disponibilité, l'intégrité et l'authentification). L'autre catégorie comprend des objectifs secondaires propres aux spécificités des RCSF (l'auto-organisation, la localisation sécurisée, le temps de synchronisation et la résistance aux attaques) [2] [3].

### 1. La confidentialité (Confidentiality)

Elle consiste à s'assurer que lors de la communication entre deux nœuds X et Y, personne ne pourra interférer afin de récupérer les données transférées. De nos jours, La plupart des mécanismes de cryptage peuvent assurer une confidentialité optimale à base des clés de chiffrement de grande taille. Cependant, ces algorithmes sont gourmands en termes de mémoire et de calcul, ce qui rend leurs adaptations aux contraintes énergétiques des RCSF trop compliqué.

### 2. L'intégrité (Integrity)

L'intégrité des données garantit que le message n'a pas été modifié lors de la propagation. Les données devraient être accessibles uniquement pour utilisateurs autorisés. Afin de garantir l'intégrité des données, il faut associer à ces données une redondance codée qui atteste la validité de ce qui est reçu avec une plus grande probabilité que celle d'un CRC (*Cyclic Redundancy Check*, comme le fait CRC pour détecter les erreurs de transmissions).

### 3. L'authentification d'un nœud (Node Authentication)

Permet à un nœud de s'assurer et vérifier l'identité de son destinataire. Un nœud compromis peut lancer une attaque Sybil par exemple dans laquelle il peut envoyer des données sous plusieurs fausses identités afin de corrompre les données agrégées. L'authentification d'un nœud pourra être assurée par que par l'adoption des systèmes de chiffrement symétrique et asymétriques.

### 4. La disponibilité

Afin de garantir la survie des services de réseau contre les attaques par déni de service. Ce service désigne la capacité du réseau à assurer ses services pour maintenir son bon fonctionnement en garantissant aux parties communicantes la présence et l'utilisation de l'information au moment souhaite.

### 5. La fraîcheur de données

Il est également nécessaire que les données soient récentes et que aucun ancien message n'a été redéployé. Pour atteindre ce but un compteur de temps devra être ajouté au paquet.

### 6. L'auto-organisation

Selon différentes situations, un réseau de capteurs sans fil exige que chaque nœud de capteur soit suffisamment indépendant et souple pour assurer la fonction d'auto-organisation et auto-guérison. Dans le réseau de capteur il n'y a aucune infrastructure fixe disponible pour la gestion de réseau, ce qui rend la sécurité des réseaux de capteurs sans fil plus difficile.

### 7. Le temps de synchronisation (Time Synchronisation)

La plupart des applications de réseaux de capteurs dépend en quelque forme de temps de synchronisation. Un nœud peut également exiger un calcule de délai de bout en bout d'un paquet qui se déplace entre deux à deux extrémités.

### 8. La Localisation sécurisée

L'utilité d'un réseau de capteurs dépend principalement de sa capacité à localiser automatiquement et avec précision chaque capteur dans le réseau. Afin de détecter un problème, un réseau de capteurs aura besoin d'informations de localisation précise. Un attaquant peut facilement exploiter cette situation et peut facilement manipuler les informations de localisation non-sécurisé par la force de faux signaux.

## III. Attaques et contremesures

Les différentes caractéristiques des réseaux de capteurs sans fil (énergie limitée, faible puissance de calcul, et l'utilisation des ondes radio, etc. ...) les exposent à de nombreuses menaces de sécurité. La vulnérabilité inhérente des réseaux de capteurs, qui sont généralement plus liés à des menaces de sécurité physiques, introduit de nouveaux et véritables challenges de sécurité. Le déni de service, et les attaques d'usurpation d'identité sont plus faciles à mener dans ce type de réseaux sans fil et sans infrastructure.
Nous décrivons dans cette partie les principaux attaques et menaces de sécurité des RCSF.

### 1. L'attaque Jamming

C'est une attaque de type Déni de Service (DoS) dont le but est de perturber ou de bloquer la communication. L'attaquant met en place une antenne afin d'émettre sur la même bade de fréquence pour empêcher les nœuds de communiquer sur ces fréquences et provoquer des interférences radio. Il existe différentes stratégies pour provoquer une attaque de type jamming. En émettant sans interruption un signal radio (constant jamming). Cette technique engendre une consommation d'énergie trop élevé. La deuxième technique, en émettant régulièrement à intervalle fixe ou de façon aléatoire sur le canal (deceptive jamming Vs random jamming) afin de réserver son énergie. La dernière technique consiste à ne pas émettre un signal que si le canal est actif (reactive jamming) [2].
Il existe plusieurs stratégies simples qui permettent de limiter et contrer le déni-de-service causé par l'attaque *Jamming*. Par exemple, si un nœud reçoit successivement des signaux et arrive à identifier qu'il s'agit d'un *Jamming*, il pourra passer en mode veille pour un laps de temps (ce qui provoquera une réaction au niveau du routage des paquets). Il se réveillera de temps en temps pour voir si l'attaque est toujours active [4]. Autre technique consiste à isoler la zone infectée afin de contourner les nœuds malicieux.

### 2. L'attaque Sélective Forwarding

Dans les réseaux de capteurs sans fil, chaque capteur participe au routage des données de ses capteurs voisins, et l'attaque « Selective Forwarding » exploite cette dépendance, afin de provoquer un déni de service (DoS), par exemple en négligeant de renvoyer un message, de la station de base ou à partir d'un autre capteur.
Pour se défendre de l'attaque "Selective Forwarding", il est possible d'utiliser l'approche "diversity coding" qui peut atténuer les effets de l'attaque en envoyant les messages codifiés sur plusieurs chemins.

### 3. Capture physique de nœuds

Une des caractéristiques des RCSF est le déploiement de leurs nœuds dans des environnements où ils sont censés être autonomes. Une fois déployés, les nœuds établissent les liens entre eux et constituent les chemins de routage sans intervention extérieure. Ces propriétés facilitent le rôle des adversaires surtout dans les régions d'accès facile où les nœuds peuvent être attrapés ou physiquement manipulés.

L'adversaire doit posséder les équipements techniques et la connaissance suffisante pour mettre en œuvre ce genre d'attaque. Il faut aussi prendre en considération les menaces liées à son intention comme le vol ou la destruction des nœuds capturés.

Nous pouvons classer les attaques Tampering et Node Compromise dans la même catégorie car elles ont besoin d'un accès physique aux nœuds. Nous détaillons ces deux attaques dans la suite.

### 4. L'attaque Tampering

Cette attaque consiste à capturer physiquement au nœud afin d'extraire et exploiter toutes les informations présentes comme les clés de chiffrement, afin de les réutilisés ultérieurement pour lancer d'autres types d'attaques.

Plusieurs travaux ont été proposés pour contrer ce type d'attaque. Les auteurs dans [5] et [6] suggèrent des modèles où les clés pré-distribuées sont supprimés après leur utilisation dans l'établissement d'un canal de communication sûr. Il faut alors vérifier que le temps nécessaire pour l'établissement, la génération et la suppression des clés soit plus petit que celui de la capture physique et de l'extraction des informations. Une telle attaque exige l'utilisation d'un protocole de gestion de clés pour protéger les nœuds.

### 5. L'attaque Sybil

Un nœud malveillant peut prétendre avoir de multiples identités en utilisant les identités des nœuds ciblés par l'attaque. L'attaque Sybil est le nom porté par ce genre de menaces. Les nœuds ciblés par l'attaque sont connus sous le nom de « Sybil nodes ». Cette attaque est localisée entre la couche liaison et la couche réseau. Elle vise à dégrader l'intégrité des données, le niveau de sécurité et l'utilisation des ressources [7].

L'authentification des nœuds est la solution optimale pour contrer ce genre d'attaques.

### 6. L'attaque Sinkhole

Cette attaque fait partie de l'ensemble des attaques de type déni-de-service. Dans cette attaque, le nœud malveillant agit comme une station de base en attirant les paquets vers lui en informant les autres nœuds du réseau qu'il a le chemin le plus cours vers la station de base « Meilleure métrique », ce qui pousse les nœuds a l'acheminé toute les information de routage. L'attaque Sinkhole est parfois assimilée à une attaque de type Black Hole (i.e., le trou noir) où un nœud malveillant mis en place un trou noir dans le réseau afin de détruire les paquets arrivant vers lui [8]. Cependant, l'attaque Sinkhole est plus complexe. Afin de préserver l'énergie des nœuds, les protocoles de routage dans les RCSF sont souvent basés sur des objectifs comme l'exploitation des chemins avec un minimum de sauts ou un minimum de délai sur les liaisons de bout-en-bout. Les attaques comme Sinkhole exploitent ces propriétés en proposant de faux chemins optimaux aux nœuds du réseau, c'est par ce fait qu'elles attirent les paquets vers un nœud qui va les absorber.

Dans les travaux [9] et [10], les auteurs ont proposés des protocoles de routage basés sur l'utilisation de la position géographique des nœuds. Ce type de protocoles parait comme solution optimale pour l'attaque Sinkhole et Blackhole, puisqu'il exploite la position du nœud comme source pour déterminer le chemin de routage. Les protocoles de cryptage à clé publique permettent eux aussi de lutter contre ce genre d'attaques.

### 7. L'attaque trou de ver (Wormhole)

Dans ce genre d'attaque un attaquant enregistre paquet à un endroit, un tunnel vers un autre emplacement dans le réseau, puis les retransmettre dans le réseau. La figure 2 illustre un exemple d'attaque de type Wormhole. L'attaquant forme un lien ou tunnel de faible latence vers un autre nœud malicieux dans le réseau, et tous les paquets seront acheminés par le chemin formé par $X_1$ et $X_2$.

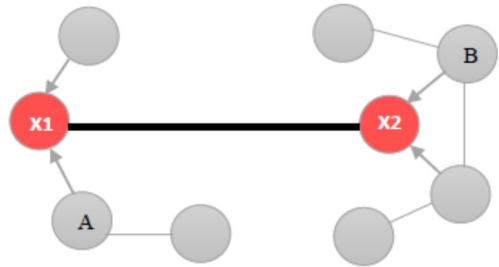

Figue 2. L'attaque Trou de Ver (Wormhole)

Plusieurs approches ont été proposées pour contrer cette attaque. On peut classes ces solutions ont deux classes : approches centralisées ou les données des nœuds voisins d'un saut seront acheminées à une entité centrale qui prend en charge la construction du schéma de routage réseau et par la suite pourra détecter les chemins inconsistants qui indiques des liens Wormhole. La deuxième technique décentralisée ne nécessite pas la présence d'une entité de gestion centrale. Chaque nœud forme son propre modèle de voisins en se basant sur un système de positionnement géographique. D'autres solutions exigent une synchronisation des horloges du réseau.

## 8. L'attaque Hello flood

Dans l'attaque Hello Flood, l'attaquant utilise des paquets hello comme une arme pour convaincre les capteurs dans le réseau. L'attaquant envoie un paquet de type Hello d'un nœud à un autre avec plus d'énergie. Un attaquant ayant le pouvoir de transmettre utilise un signal radio trop puissant à longue portée pour envoyer des paquets de ce genre et inonder une partie du réseau tout en provoquant de fausses listes de voisins. Les capteurs reçoivent ces paquets sont donc convaincu que l'adversaire est leur voisin. En conséquence, les nœuds victimes essaient de passer par l'attaquant, car ils savent qu'il est de leur voisin tout en envoyant l'information à la station de base (BS), comme le montre la figure 3.

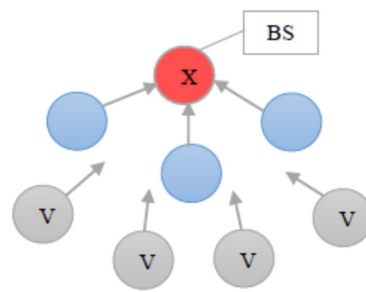

Fig. 3. L'attaque Hello flood

Les protocoles de sécurité basés sur l'identification des nœuds peuvent être une solution pour vérifier leurs identités, et par conséquent empêcher un attaquant à produire une attaque de type Hello flood.

## IV. La cryptographie dans les réseaux RCSF

Les RCSF sont exposés face à divers menaces de la sécurité à cause de l'utilisation des ondes radio comme support de transmission, ce qui nécessite l'adoption des méthodes spécifiques pour pouvoir sécuriser ce type de réseaux. Les moyens de protection traditionnels comme les protocoles de chiffrement utilisés dans les réseaux filaires ou wifi sont suffisamment sécurisés pour être utilisés dans les RCSF mais l'effort d'adaptation représente un véritable challenge. L'adaptation de n'importe quel mécanisme de sécurité aux exigences des RSCF doit respecter un certain nombre de conditions. La rapidité de l'échange de clés et des informations qui concernent leur création puisque les communications passent via les signaux radio et risquent d'être interceptées à tout moment. La nature du déploiement des RCSF pour certaines applications rend les nœuds vulnérables à une capture physique (la perte de ces nœuds ne doit pas affecter la sécurité des communications de tout le réseau). L'architecture proposée pour une application donnée doit trouver un compromis entre le niveau de sécurité demandé et le coût total de consommation de ressources [11]. Par exemple, l'utilisation d'une seule clé de chiffrement pour tout le réseau permet de diminuer l'occupation mémoire, de supprimer le calcul de clés dérivées et ainsi d'économiser de l'énergie. Cependant, le chiffrement à l'aide de cette clé permet d'assurer la confidentialité des communications pour une courte durée seulement. Un adversaire peut extraire la clé à partir d'un nœud capturé physiquement et déchiffrer toutes les informations qui circulent dans le réseau. D'où l'intérêt d'avoir une architecture suffisamment sécurisée tout en respectant la limitation des ressources des nœuds. Il existe dans la littérature plusieurs mécanismes de chiffrement pour les RCSF basées sur le système de chiffrement symétrique, le système asymétrique et le système hybride (qui utilisent les deux autres à la fois). Nous avons fait une classification qui tient compte des différents types de solutions.

### 1. Méthodes et protocoles de cryptage pour les RCSF

La plupart des méthodes basées sur des systèmes symétriques, asymétriques ou hybrides résoudre le problème de l'établissement de clé par une phase de pré distribution. Le pré distribution de clés de cryptage dans un RCSF est le fait de stocker ces clés dans les nœuds de mémoire avant le déploiement. Dans la littérature, on trouve plusieurs classifications des systèmes de gestion des clés cryptographiques, tels que les papiers à [12] - [14]. Certaines méthodes de classification sont basés sur le partage du clavier entre deux nœuds (Pair-Wise) ou plusieurs nœuds (Groupe-sage), et d'autres comptent sur l'exploitation des probabilités, l'analyse combinatoire, etc. Nous avons choisi de faire une classification, qui inclut tous les modèles de gestion et de distribution clés en deux grandes familles, schémas asymétriques et symétriques. La figure 4 illustre cette classification. Dans ce qui suit, nous allons détailler les principaux modèles de la littérature.

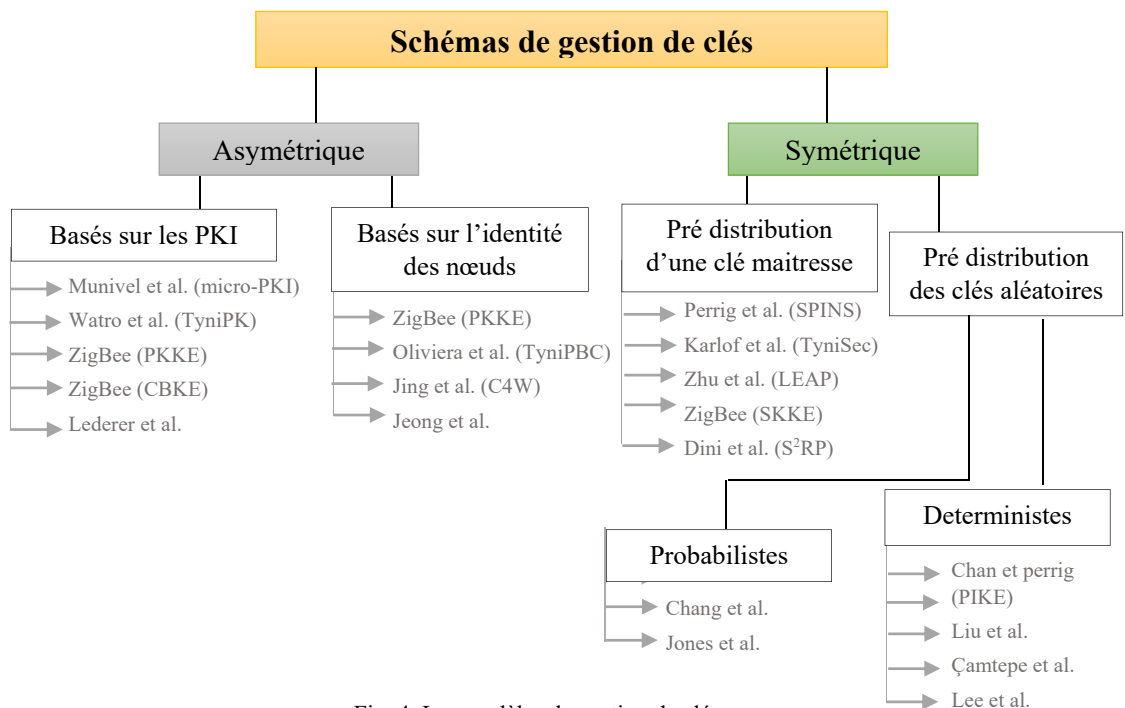

Fig. 4. Les modèles de gestion de clés

### 1. Schémas de gestion de clés symétriques

Les schémas dans cette catégorie utilisent des mécanismes symétriques afin d'établir une clé commune entre deux nœuds dans un RCSF. Ceci est réalisé en trois étapes :

- Pré distribution des clés *(Key predistribution)* : les clés stockées dans la mémoire avant le déploiement constituent le porte-clés (Key ring). S'il existe une clé commune entre deux nœuds, ils peuvent créer une connexion sécurisée entre eux.
- Découverte de la clé partagée *(Shared-Key decouvert)* : Après le déploiement, le protocole de communication est responsable de la découverte de la clé commune entre deux nœuds voisins.
- Création et établissement de chemin sécurisé par des clés *(Path-key establishement)* : s'il n'y a pas de clé commune entre deux nœuds qui souhaitent communiquer, il faut alors trouver un chemin sécurisé entre eux. Ce chemin passe par un ensemble de nœuds qui contient déjà des liens sécurisés. Une fois le chemin établi, les deux nœuds peuvent l'exploiter pour sécuriser les communications.

Nous présentons ci-après les célèbres schémas symétriques proposés pour sécuriser les RCSF :

### 1.1 SPINS

SPINS est une suite de blocs de construction de sécurité proposé par Perig et al. [15]. Il est optimisé pour les environnements à ressources limités. SPINS dispose de deux blocs de constructions sécurisées : SNEP et μTESLA. SNEP utilise un compteur partagé entre les deux parties communicantes et applique le compteur dans le calcul du code d'authentification de message (MAC) pour assurer la confidentialité et l'authentification des données. De plus, le protocole a aussi une faible surcharge, puisque le compteur est maintenu à chaque point d'extrémité et le protocole ne fait qu'ajouter 8 octets par message. μTESLA assure une diffusion authentifié. SPINS fournit une méthode optimale pour étendre la confiance établie entre les nœuds et la station de base aux liens établis entre les nœuds directement.

Le mécanisme de fonctionnement du protocole SPINS consiste à stocker une clé secrète maîtresse (M) dans le nœud capteur et aussi dans sa station de base. Après le déploiement, un nœud X et une station de base Y dérivent chacun deux clés secrètes à partir de la clé maîtresse ($M_{XY} = M_{YX}$). Le nœud X obtient $K_{XY}$ et $K'_{XY}$, et le nœud Y obtient $K_{YX}$ et $K'_{YX}$. La première clé est utilisée pour le chiffrement et la deuxième est dédiée au calcul du code d'authentification des messages (MAC). La dérivation se fait à l'aide d'une fonction pseudo-aléatoire F. Puisque les deux nœuds partagent la même clé maîtresse et la même fonction de dérivation, chaque nœud pourra dériver les clés secrètes de l'autre pour pouvoir déchiffrer et vérifier les messages reçus. Au lieu d'envoyer les compteurs avec les messages, X et Y partagent deux compteurs (un pour chaque direction de la communication). Ils sont synchronisés à l'aide d'un protocole d'échange de compteurs, $Compt_X$ et $Compt_Y$ sont les compteurs respectifs de X et Y. Les deux compteurs ne sont pas secrets, alors X peut envoyer son compteur à Y en clair. À la réception, Y envoie son compteur accompagné du code d'authentification (MAC) de la combinaison des deux compteurs [2]. Ainsi, A pourra vérifier l'intégrité de $Compt_Y$. Il envoie ensuite le code d'authentification de la combinaison des deux compteurs à B afin qu'il puisse vérifier l'intégrité de $Compt_X$. La figure 5 montre le protocole d'échange de compteurs.

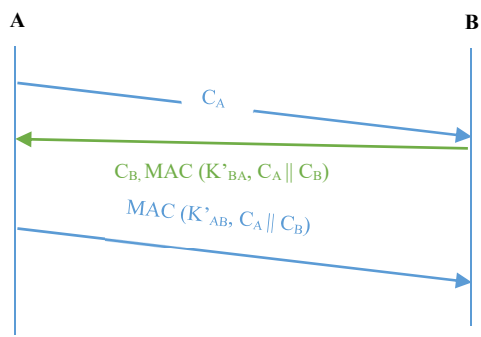

Figure 5. Protocole d'échange de compteurs

En remplaçant l'envoi des compteurs dans chaque message par un calcul de MAC (code) et une synchronisation de compteur, la méthode de partage de compteur économise de l'énergie. On peut noter que cette méthode ne pourra être bénéfique que si les deux nœuds communicants ont beaucoup de données à échanger. De plus, chaque communication a besoin de deux compteurs synchronisés (un par direction), et ceci pourrait surcharger le calcul et le stockage mémoire des nœuds. Cependant, SPINS semble avoir encore quelques problèmes sous-jacents comme suit :
- Il ne considère pas la possibilité d'attaque DOS.
- Grâce à utiliser le système de pré-distribution de clé par paires dans le protocole de routage de la sécurité, SPINS repose sur la station de base de façon excessive.
- SPINS ne considère pas la mise à jour de la clé de communication. Il doit y avoir un mécanisme de mise à jour de clés pour avoir une sécurité profonde.
- SPINS ne peut pas résoudre le problème de nœud compromis.

*1.2 LEAP*

LEAP (Localized Encryption and Authentication Protocol) est un protocole de gestion de clés pour les réseaux de capteurs qui est conçu pour supporter le traitement interne en réseau *"In network processing"* avec l'objectif premier est de limiter l'impact de sécurité d'un nœud compromise dans le réseau sur son voisinage immédiat [5]. L'idée de la conception du protocole LEAP a été motivée après avoir observé que les différents types de messages qui sont échangés entre les nœuds de capteurs ont des exigences différentes de la sécurité. LEAP supporte la création et l'établissement de quatre types de clés : clé Individuelle, clé par-paire, clé de groupe et clé globale. L'utilisation de ces quatre clés permet de minimiser l'implication de la station de base dans le processus de gestion de clés, ce qui réduit en conséquence la consommation d'énergie et le trafic. Le protocole LEAP repose sur l'hypothèse suivante : un attaquant nécessite un temps minimal Tmin pour compromettre un nœud et copier le contenu de la mémoire. LEAP essaye d'exploiter ce temps pour permettre à deux nœuds voisins d'établir une clé symétrique, à partir de la clé Kin pré-chargée sur chaque capteur avant le déploiement, et de supprimer cette dernière de la mémoire du nœud compromis en un temps T < Tmin.

*1.3. Tinysec*

Karlof et al. [16] proposent le Protocole TinySec, la première mise en œuvre complète d'une architecture sécurisée à la couche liaison de données pour les RCSF. Cette mise en œuvre prend en charge deux options de sécurité : un message d'authentification avec cryptage de données (TinySec-EA) et l'authentification des messages sans cryptage des données (TinySec-Auth). Comme SPINS, TinySec utilise des algorithmes standards de cryptographie pour assurer la confidentialité et l'intégrité. Les auteurs de Tinysec trouvent que l'algorithme Skipjack [17] est plus approprié pour les RCSF que RC5 (algorithme utilisé par SPINS). En effet, les évaluations de TinySec ont montré que RC5 a besoin d'un calcul préalable à l'aide de 104 octets de RAM. TinySec utilise le mode de chiffrement CBC (Cipher Block Chaining) au lieu du CTR (utilisé par SPINS). En effet, le CTR fournira pour plusieurs cryptages de paquets les mêmes nombres aléatoires. Ces nombres sont utilisés principalement dans la production de séquences de clés de chiffrement, leur répétition peut affaiblir le niveau de cette solution de sécurité et en conséquence permettre aux intrus de découvrir le contenu des messages. TinySec est une mise en œuvre plutôt qu'une proposition de distribution de clé. Son objectif est de compléter une méthode de distribution de clé adapté au réseau déployé. Deux nœuds ont besoin de deux clé symétrique partagée pour communiquer. La première est utilisée pour chiffrer les messages et la seconde pour le calcul de MAC (code) des messages.

2. **Schémas de gestion de clés asymétriques**

Les schémas dans cette catégorie utilisent les mécanismes des systèmes asymétriques afin d'établir une clé commune entre deux nœuds ou un groupe de nœuds d'un RCSF.

*2.1. micro-PKI*

Munivel et al. [18] proposent un protocole pour les RCSF nommé micro-PKI (Micro Public Key Infrastructure), une version simplifiée des PKI conventionnelles. La station de base a une clé publique et une autre privée. La clé publique est utilisée par les nœuds de réseau pour authentifier la station de base, et la clé privée est utilisée par la station de base pour décrypter les données envoyées par les nœuds. Avant le déploiement, la clé publique de la station de base est stockée dans tous les nœuds. Les auteurs incluent dans leur régime deux types d'authentification (Handshake). Le premier type d'authentification se produit entre un nœud de réseau et la station de base. Le nœud génère une clé symétrique de session et crypte avec la clé publique de la station de base. Pour assurer l'intégrité des messages échangés, les auteurs proposent d'intégrer à chaque message un MAC (code) en utilisant la même clé de cryptage du message. Pour les nouveaux nœuds qui souhaitent rejoindre le réseau, ils stockent simplement dans ces nœuds, la clé publique de la station de base avant le déploiement.

*2.2. TinyPK*

Watro et al. [19] ont proposé une méthode appelée TinyPK basée sur l'utilisation de clés publiques et le principe de Diffie-Hellman pour établir une clé secrète entre deux nœuds dans un RCSF. TinyPK utilise une autorité de confiance pour signer les clés publiques de nœuds. La clé CA est pré distribuée à tous les nœuds avant le déploiement afin qu'ils puissent vérifier les clés vde ces voisins après le déploiement. Le choix de l'algorithme RSA pour le cryptage implique une grande consommation de temps et d'énergie des nœuds. Ainsi, les opérations de base peuvent prendre douzaine de secondes, ce qui permettra de réduire la durée de vie du réseau ainsi que l'impact sur la réactivité.

*2.3. PKKE & CBKE*

Les protocoles PKKE et CBKE proposés par Zigbee utilisant l'identité de nœuds dans leur méthode d'établissement de clé. Le but est d'utiliser ces identités pour créer une seule clé partagée entre chaque paire de nœuds dans un réseau. Toutefois, la création de la clé partagée est réalisée avec les interactions entre les deux nœuds. Cela veut dire, les méthodes nécessitent l'envoi et la réception de plusieurs messages sur les deux côtés avant la mise en place de la clé. Pour économiser l'énergie des nœuds souhaitent partager un message secret et les nœuds intermédiaires, plusieurs méthodes ont été proposées pour éliminer ces interactions. Ces méthodes sont connues dans le domaine de la cryptographie comme ID-NIKDS [20] (*Identity-Based Non-Interactive Key Distribution Scheme*).

*2.4. C4W*

Jing et al. [21] ont proposé une méthode appelée C4W basée sur l'utilisation de l'identité des nœuds pour calculer les clés publiques. Les nœuds eux-mêmes sont en mesure de calculer les clés publiques des autres nœuds en utilisant leurs identités. Que pourrait remplacer le rôle d'un certificat. Avant le déploiement, les nœuds et la station de base sont chargés avec leurs propres clés (clé privée / publique ECC) et l'information du public sur les nœuds du réseau. La méthode C4W utilise le principe de l'échange de clés Diffie-Hellman pour créer une seule clé partagée entre deux nœuds, sans l'aide de certificats

### 3. Synthèse

Les diagrammes SPINS et LEAP utilisent des clés maîtresses dans l'établissement de clés. Cela réduit le stockage de clés dans la mémoire de nœuds. Cependant, la résistance aux attaques est faible. Étant donné que la clé principale peut être compromise à tout moment, les clés établies après le déploiement en utilisant cette clé peuvent être compromises aussi. Les systèmes symétrique sont les plus appropriés et parmi les plus rapides en termes de calcul. Notez que les diagrammes symétriques sont coûteux en opérations (si elles existent) de renouvellement et de révocation des clés car ils utilisent des clés secrètes afin d'échanger d'autres clés secrètes. Le problème est plus simple dans les schémas asymétriques puisque les clés publiques n'ont pas besoin d'être secrètes.
Nous pouvons observer que le schéma TinyPBC est le plus approprié parmi les modèles asymétriques. Il assure une grande résistance aux attaques les plus connues dans le RCSF. Le fait d'utiliser l'accouplement afin d'établir une clé unique partagée entre deux nœuds a permis de réduire le besoin de grande capacité de stockage en mémoire. En outre, la création de cette clé est effectuée sans interaction entre les nœuds, ce qui permet d'économiser le temps de calcul et de l'énergie consommée en raison de ces interactions. Les diagrammes basés sur le principe des certificats PKI et restent le plus cher dans le calcul et la consommation d'énergie.

La comparaison entre les schémas symétrique et asymétrique peut différer en fonction du niveau de sécurité souhaité dans le réseau. Nous notons que les diagrammes symétriques peuvent être choisis pour leur rapidité et les schémas asymétriques pour leur résistance contre les attaques.

## V. Un nouveau prototype de chiffrement pour les RCSF

Dans le chapitre précédent, on a tenté de donner une description générale sur l'état de l'art des algorithmes de cryptage proposés dans le monde des RCSF. On peut en conclure que les nœuds de réseaux de capteurs sont généralement déployés dans un environnement hostile sans surveillance comme par exemple un champ de bataille. Par conséquent, il est extrêmement important pour les applications critiques des réseaux de capteurs d'avoir un mécanisme de sécurité qui assure l'authentification et la confidentialité. Il est particulièrement difficile d'assurer une sécurité optimale dans ce type de réseau en raison de la limitation des ressources des nœuds. En gardant en vue ces limitations, différentes études des protocoles de cryptage ont été mené dans le but de proposer un algorithme de cryptage symétrique pour la gestion des clés. Les deux algorithmes de base pour les schémas proposés sont "Leap+" et "l'amélioration de la négociation des clés". Une approche simple et pratique pour l'amorçage de clés secrètes dans les réseaux de capteurs est le pré déploiement de ces clés. Le degré de partage de clés entre les nœuds dans le système est un facteur de conception très important pour les protocoles de sécurité basés sur des clés symétriques. Il peut y avoir deux extrémités pour le chiffrement des données et l'authentification, d'un côté il pourrait avoir l'ensemble de clés du réseau et à l'autre extrême il pourrait y avoir une approche de répartition de clés par laquelle toute la communication sécurisée est basée sur des clés qui sont partagés par-paire entre deux nœuds. Le premier cas présente les coûts de stockage les plus bas et il est efficace en termes d'énergie, car il n'a pas besoin de communication entre les nœuds pour établir des clés supplémentaires. L'inconvénient de sécurité associé à ce schéma est que le compromis d'un seul nœud révèle la clé globale et donc l'ensemble du réseau devient à risque. Le deuxième cas est idéal en termes de sécurité, car le compromis de nœud ne révèle pas toutes les clés qui sont utilisés par d'autres nœuds du réseau.

La question unique qui doit être considéré dans un réseau de capteur avant de choisir une approche de partage de clé est de prévoir son impact sur l'efficacité du traitement en réseau. L'algorithme proposé supporte le traitement en réseau "in-network processing'', et fournit également des propriétés de sécurité semblables à celles des schémas de partage de clés par-paire. Similaire à LEAP+ et autres protocoles, la solution proposée est également basée sur le fait que les différents types de messages échangés entre les nœuds de capteurs ont différentes exigences de sécurité, ce qui nous a permis de déduire qu'un mécanisme de chiffrement unique ne convient pas à répondre à ces exigences de sécurité différentes [5]. A l'instar de LEAP+, l'algorithme proposé supporte l'établissement de quatre types de clés différentes :

- Clé individuelle (Individual Key) : Commune à la station de base,
- Clé par paires (Parwise Key) : Partagée avec un autre nœud de capteur,
- Clé de Cluster (Cluster Key) : Partagée Avec plusieurs nœuds voisins,
- Clé Global (Global Key) : partagée par tous les nœuds dans le réseau.

## 1. Hypothèses en termes de réseau et de sécurité

Les hypothèses suivantes ont été prises en considération pendant l'étude et la conception du protocole

- Un réseau de capteurs statique où les nœuds ne sont pas mobiles,
- La station de base agit en tant que contrôleur,
- La puissance énergétique fournie à la station de base pour pouvoir durer longtemps,
- Tous les nœuds sont égaux en capacités de calcul et de communication,
- Chaque nœud dispose d'espace suffisant pour stocker les clés de chiffrement,
- L'installation des nœuds peut être faite de deux manières, soit par une installation physique ou une diffusion aérienne,
- À l'avance les nœuds voisins d'un nœud de capteur ne sont pas connus,
- Toutes les informations stockées dans un nœud capteur deviennent connu à l'attaquant si elle est compromise,
- Les attaques de la couche physique et la couche contrôle d'accès au support ne sont pas considérés.

## 2. Aperçu de base de l'algorithme

Les nœuds échange différents types de paquets basés sur différents critères :
- Paquets de contrôle Vs paquets de données
- Paquets de broadcast Vs Paquets Unicast
- Requêtes ou commandes Vs Lectures des capteurs et ainsi de suite

Les exigences de sécurité pour chaque type de paquets sont différentes. L'authentification est requise pour tous les types de paquets alors que la confidentialité est nécessaire pour certains types de paquets. Comme il a été expliqué auparavant, un mécanisme de chiffrement unique ne convient pas à toutes les communications sécurisées nécessaires dans les réseaux de capteurs. L'algorithme proposé à l'instar de LEAP supporte quatre types de clés pour chaque nœud :

*2.1 Clé individuelle (Individual Key)*

Pour avoir une communication sécurisée entre le nœud et la station de base cette clé est utilisée. Chaque nœud dans le réseau possède sa propre clé individuelle. La clé individuelle est également importante pour calculer le code d'authentification de message (MAC) si le message doit être vérifié par la station de base. Cela peut également être utilisé pour envoyer une alerte à la station de base en cas de comportement anormal observé. La station de base peut utiliser la clé individuelle pour crypter les informations sensibles telles que des clés partagées ou des instructions spéciales à un nœud précis. Il est important de mentionner que la clé individuelle est pré chargé dans le réseau avant le déploiement. La clé individuelle est générée comme suit :

$$Iku = FKM(u) \qquad (1)$$

Où *Iku* est la clé individuelle, *F* est une fonction pseudo-aléatoire, *Km* est la clé maitresse, et u est un nœud pour lequel nous voulons trouver la clé individuelle.

Il est également possible de conserver le stockage nécessaire pour maintenir toutes les clés individuelles. La station de base (le Contrôleur) calcule la clé individuelle à la volée lorsque cela est nécessaire.

*2.2 Établissement des clés-paire (Pairwise Key)*

L'étape la plus importante est d'avoir la clé paire entre les nœuds. Il est très primordial du point de vue de la sécurité que si la clé est compromise son effet est localisé. La clé paire est partagée entre les voisins d'un saut. Un nœud communique avec son voisin immédiat par la clé paire. Les hypothèses importantes ci-après sont prises en compte pour établir les clés paires.

- Le nœud ne connaît pas la clé paire de son voisin avant le déploiement. La clé paire est créée après le déploiement
- Les nœuds du réseau sont des nœuds stationnaires
- Un nœud qui est ajouté au réseau découvre la majeure partie de ses voisins, au moment du déploiement.

La clé paire est générée en suivant ces étapes :

### 2.2.1 Key pré répartition

Une clé initiale générée par le contrôleur (Station de base) est généré pour chaque nœud. Chaque nœud dérive par la suite sa clé maîtresse ainsi :

$$Ku = fKin(u) \qquad (2)$$

Où *Ku* est la clé maitraisse générée par le nœud *u*, *f* est une fonction pseudo-aléatoire, *Kin* est la clé initiale, et *u* est n'importe quel nœud pour laquelle nous voulons déruver la clé principale.

### 2.2.2 Neighbor Discovery

Chaque nœud tente de trouver son voisin par la diffusion d'un message HELLO. Ce message de contient l'identifiant (ID) du nœud. Aussi une temporisation est débutée et qui expire après un certain temps Tmin. Après, ce nœud *u* attend n'importe quell nœuf *v* qui répond à ce message Hello avec un message ACK ayant l'id du nœud *v*. Ack du voisin est authentifié en utilisant la clé maîtraisse *kv*. La clé maitraisse est calculée ainsi :

$$Kv = fKin(v) \qquad (3)$$

Où *Kv* est la clé principale du nœud *v*, f est une fonction pseudo-aléatoire, *Kin* est la clé initiale, et v est n'importe quel nœud que nous souhaitons trouver sa clé maitresse.

### 2.2.3 Etablissement de la clé par-paire

Tout nœud u and v dans le réseau calcule la clé par-paire ainsi :

$$Kuv = FKV(u) \qquad (4)$$

Où *kuv* est la clé par paires entre le nœud *u* et *v*, f est une fonction pseudo-aléatoire, *kv* est la clé principale de nœud, et *u* est l'ID de nœud d'un nœud *u*.

### 2.2.4 Effacement de clés

Lorsque le Timer expire après Tmin, le nœud *u* efface *Kin* et toutes les clés maîtresses de son voisin qui ont été calculé pendant la phase de découverte de voisinage.

*2.3 Définir les clés de cluster*

La clé de cluster est établie entre un nœud et tous ses voisins. Avec l'utilisation de clés de cluster, un nœud chiffre les messages de broadcast. Pour établir la clé de cluster tout nœud u génère une clé aléatoire, puis crypte cette clé aléatoire avec la clé par-paire déjà générée. Ensuite, la clé de cluster générée est transmise à chaque voisin.

*2.4 Création de la clé Globale*

Une clé partagée par la station de base et chaque nœud est la clé globale. Il est utilisé lorsque la station de base (contrôleur) veut générer un message confidentiel.

### 4. Notre Contribution

### 1. Premier modèle

L'hypothèse critique que LEAP+ a considérée est que dans Tmin, un nœud ne peut pas être compromis. Cette idée semble pratique, mais seulement dans une condition extrêmement idéale. Il est possible que Tmin soit en réalité plus grande que celle assumée. A titre d'exemple, si les nœuds sont supprimés et dispersés du réseau, les nœuds dispersés peuvent arriver dans différentes parties du réseau à différents moments, même en cas de suppression simultanée, et donc les nœuds auront besoin de temps pour reformer le réseau et échanger la clé par-paire. Prenant cet avantage, un adversaire peut observer un nœud et obtenir la clé et si la clé globale est compromise l'ensemble du réseau devient à risque. Ceci pourra déclencher des menaces très graves pour la sécurité. Pour contrer ce type de menaces, plusieurs algorithmes ont proposé différents modèles pour détecter le nœud compromis et prendre les mesures nécessaires pour le supprimer du réseau [22] [23] [24] [25] [26]. Voici quelques algorithmes disponibles dans la littérature pour détecter les nœuds compromis :

- Detecting Compromised Nodes in Wireless Sensor Network by Rick Mckenzie, Min song, Mary Mathews, sachin shetty,
- A Framework for identifying Compromised Nodes in sensor Network by Qing Zhang, Ting Yu, Peng Ning
- Malicious Node Detection in wireless sensor Networks using by Idris M.Atakli, Hongbing Hu, Yu chen
- Sensor Node Compromise Detection, The location Perspective by Hui song and liang xie

Un temps appelé Tp a été choisi et exécuter de manière itérative après une certaine période de temps pour vérifier l'existence d'un nœud compromis. Il peut être exécuté directement après l'échange de clés par-paire "Pairewise keys", et comme ça les nœuds démarrent la communication de sorte à être sûr qu'aucun des nœuds n'est compromis et que le réseau a échangé avec succès toutes les clés en toute sécurité. Les étapes pour effectuer les tâches seraient les suivantes [27] :

*1.1 Étape 1 : Vérification périodique pour la détection de nœud compromis*

Un itérative périodique « PÉRIODIQUES-CHECK (Tp) » routine a été exécuté sur chaque nœud pour vérifier si elle est attaquée ou non. La durée de "TP" est le compromis de la "complexité" et "menace de l'attaquant". Dans le premier cas Tp pourrait être augmenté de façon à minimiser la complexité du réseau en termes d'échange de paquets. Dans le deuxième cas, menace de l'attaquent, Tp pourrait être minimisé pour avoir des contrôles plus fréquents sur la compensation de nœud.

Supposons maintenant qu'un nœud est compromise exactement après la fin du "CHECK périodique", qui est "Tp + t", dans ce cas, plutôt que d'attendre pour la prochaine période (t = 2TP supposer), le nœud lui-même envoie un message "Help broadcast". La station de base reçoit le message "Help broadcast" et prend les mesures nécessaires expliqué dans la prochaine étape.

*1.2 Étape 2 : Suspension du nœud compromis*

Lors de la réception du message "Help" à partir du nœud attaqué, la station de base diffuse un message d'ALERT. Le message d'alerte contient l'id de l'émetteur du HELP (à savoir le nœud sous attaque, qui a été nommé "in_danger_id"). Nœuds recevant un message "ALERT", inspecte pour le nœud "in_danger_id" et le compare ensuite avec le la liste des identifiants présents dans sa liste de voisins "Neighbor List". S'il trouve une correspondance, il supprime la clé par-paire déjà établie avec ce nœud. Si aucune correspondance n'est retrouvée, le nœud maintient cet id pour un laps de temps spécifié, et chaque fois qu'il initie le processus d'échange de clé par-paire, il compare cette clé afin de ne pas établir une clé par-paire avec un nœud qui est infecté. Dans la figure 6, nous présentons le modèle proposé pour la détection du nœud compromis.

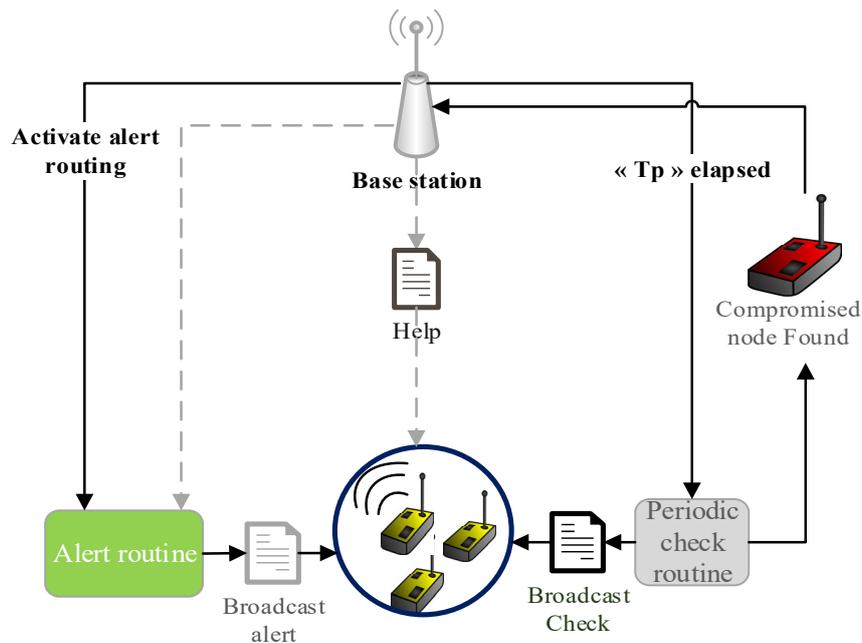

Figure 6. 1er modèle pour la détection du nœud compromis

## V. Evaluation du modèle proposé

### 1. Environnement de simulation

TOSSIM est utilisé pour simuler les applications utilisant le système d'exploitation des RCSF "TinyOS". L'objectif est de parvenir à une simulation précise et évolutive de l'application TinyOS. TOSSIM est une bibliothèque TinyOS qui fonctionnent en remplaçant les composants avec les implémentations de simulation [28] [29].
Les principales exigences pour un simulateur TinyOS sont :

a) Bridging

Le simulateur doit travailler comme pont entre l'algorithme et l'implémentation. Il doit permettre au développeur de vérifier et de tester le code qui sera ensuite implémenté sur du matériel réel.

b) L'exhaustivité (completeness)

Exhaustivité en terme de couvrir le maximum des interactions du système possible et également capturer précisément le comportement à une large échelle de niveaux.

c) Fidelity

Le simulateur doit avoir la capacité de capturer le comportement du réseau. Il est important de vérifier les interactions de synchronisation entre les nœuds à la fois pour l'évaluation et les tests.

d) Évolutivité

Le simulateur doit avoir la capacité de gérer de grands réseaux de milliers de nœuds dans une large échelle de configurations.

### 2. Configuration d'un réseau Dans TOSSIM

Pour simuler le comportement du réseau sur Tossim, vous devez spécifier une topologie de réseau. Il est facile de modifier la simulation de radio sous-jacente dans Tossim en raison de sa nature structurée. Ici, nous allons expliquer brièvement la méthode de configuration d'un réseau. [30]
Nous fournissons non seulement un ensemble de données sur le simulateur qui décrit la force de propagation, mais aussi indique le niveau de bruit et de sensibilité du récepteur. Tossim fournit des primitives de bas niveau qui expriment un large éventail de radios et de comportement.
TOSSIM simule le bruit et les interférences RF qu'un nœud entend à la fois de sources extérieures et les autres nœuds. Il utilise l'algorithme CPM de qui est plus proche pattern matching. CPM après la prise les traces de bruit dans son entrée, il génère un modèle statistique. La Configuration CPM doit ajouter une trace de bruit, en appelant addNOiseTracceReading sur un objet de promouvoir. Après avoir nourri toute la trace de bruit que nous appelons creatNoiseModel. Dans le répertoire /lib/tossim/noice, ce sont une série de lignes de bruit.

Voici un exemple, nous allons prendre 10 lignes de meyer-heavy.txt qui est une trace de bruit prises de la bibliothèque de Meyer de l'Université Stanford. [30] [31]

```
-39
-98
-98
-98
-99
-98
-94
-98
-98
-98
```

CPM utilise une bonne partie de la RAM, par exemple, les traces complète de Meyer-heavey prend environ 10 Mo par nœud si est pris comme entrée. Nous pouvons réduire cette surcharge en utilisant une trace plus courte qui permettra également de réduire le module "fidelity" de la simulation. L'exigence minimale est au-bail de 100 entrées, cependant CPM ne fonctionnera pas à cause de ne pas avoir suffisamment de données pour la génération d'un modèle statistique. Nous pouvons facilement créer des topologies dans un fichier, puis en utilisant un script python nous pouvons les charger et de les stocker dans un objet de radio. Comme exemple, nous allons créer un fichier texte topoogy.txt qui va ressembler à ceci :

```
1 2 -54,0
2 1 -55,0
1 3 -60.0
3 1 -60.0
2 3 -64,0
55
2 3 -64,0
```

Composé de trois valeurs de la source de la destination et le gain. La première ligne signifie une émission et 2 entend à -54 dBm et ainsi de suite. Un nœud lorsqu'il transmettre un paquet, les autres nœuds l'entendent. Nous écrivons un script python complet qui permettra de simuler la transmission de paquets avec RadioCountToLedsC, nous allons l'enregistrer en fichier ourtest.py.

- **Ourtest.py**

```python
#! /usr/bin/python
from TOSSIM import *
import sys
t = Tossim([])
```

```
r = t.radio()
f = open("topo.txt", "r")
for line in f:
s = line.split()
if s:
print " ", s[0], " ", s[1], " ", s[2];
r.add(int(s[0]), int(s[1]), float(s[2]))
t.addChannel("RadioCountToLedsC", sys.stdout)
t.addChannel("Boot", sys.stdout)
noise = open("meyer-heavy.txt", "r")
for line in noise:
str1 = line.strip()
if str1:
```

56

```
val = int(str1)
for i in range(1, 4):
t.getNode(i).addNoiseTraceReading(val)
for i in range(1, 4):
print "Creating noise model for ",i;
t.getNode(i).createNoiseModel()
t.getNode(1).bootAtTime(100001);
t.getNode(2).bootAtTime(800008);
t.getNode(3).bootAtTime(1800009);
for i in range(100):
t.runNextEvent()
```

Nous pouvons le lancer en tapant ourtest.py et nous obtenons le résultat suivant

Résultat du code ci-dessus

```
1 2 -54,0
2 1 -55,0
1 3 -60.0
3 1 -60.0
2 3 -64,0
23 -64 0
DEBUG (1): Application booted.
```

```
DEBUG (1): Application booted again.
DEBUG (1): Application booted a third time.
DEBUG (2): Application booted.
DEBUG (2): Application booted again.
DEBUG (2): Application booted a third time.
DEBUG (3): Application booted.
DEBUG (3): Application booted again.
DEBUG (3): Application booted a third time.
DEBUG (1): RadioCountToLedsC: timer fired, counter is 1.
DEBUG (1): RadioCountToLedsC: packet sent.
DEBUG (2): RadioCountToLedsC: timer fired, counter is 1.
DEBUG (2): RadioCountToLedsC: packet sent.
DEBUG (3): RadioCountToLedsC: timer fired, counter is 1.
DEBUG (3): RadioCountToLedsC: packet sent.
DEBUG (1): Received packet of length 2.
DEBUG (3): Received packet of length 2.
DEBUG (2): Received packet of length 2.
DEBUG (1): RadioCountToLedsC: timer fired, counter is 2.
DEBUG (1): RadioCountToLedsC: packet sent.
DEBUG (2): RadioCountToLedsC: timer fired, counter is 2.
DEBUG (2): RadioCountToLedsC: packet sent.
DEBUG (3): RadioCountToLedsC: timer fired, counter is 2.
DEBUG (3): RadioCountToLedsC: packet sent.
DEBUG (1): Received packet of length 2.
```

**L'injection de paquets :**

Nous pouvons injecter des paquets dynamiquement dans un réseau sous le simulateur Tossim. La planification de paquet peut être effectuée à tout moment. Il est tout à fait possible qu'un nœud qui est en train de recevoir un paquet d'un autre nœud sur sa radio reçoit un autre paquet injecté [32]. RadioCountMsg.py définit le format du paquet, il convient de noter que ce paquet est contenu dans la charge des données "Data payload" d'un autre format. Supposons un noeud envoyant un RadioCountMsf sur AM, puis la structure RadioCoungMsg est mis dans le payload AM qui ressemblent à :

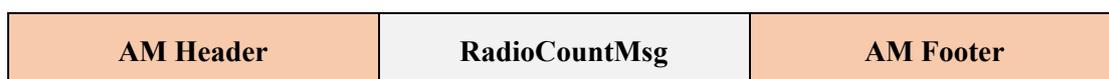

Si nous devons envoyer à travers un protocole de routage, le paquet doit être inséré dans la charge utile du routage "the routing payload" qui ressemble à

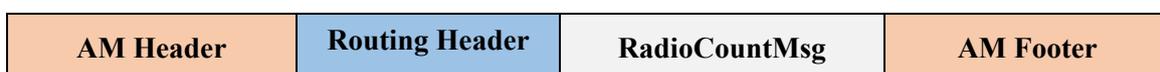

Nous allons montrer un code sous Tossim qui va initier une simulation qui configure la topologie basée sur topology.txt et qui délivre un paquet au nœud 0. Le code sera le suivant :

- **packets.py**

```python
#! /usr/bin/python
import sys
from TOSSIM import *
from RadioCountMsg import *
t = Tossim([])
m = t.mac()
r = t.radio()
t.addChannel("RadioCountToLedsC", sys.stdout)
t.addChannel("LedsC", sys.stdout)
for i in range(0, 2):
    m = t.getNode(i)
    m.bootAtTime((31 + t.ticksPerSecond() / 10) * i + 1)
f = open("topology.txt", "r")
for line in f:
    s = line.split()
    if s:
        if s[0] == "gain":
            r.add(int(s[1]), int(s[2]), float(s[3]))
noise = open("meyer-heavy.txt", "r")
for line in noise:
    s = line.strip()
    if s:
        val = int(s)
        for i in range(4):
            t.getNode(i).addNoiseTraceReading(val)
for i in range(4):
    t.getNode(i).createNoiseModel()
for i in range(60):
    t.runNextEvent()
msg = RadioCountMsg()
msg.set_counter(7)
pkt = t.newPacket()
```

```
pkt.setData(msg.data)

pkt.setType(msg.get_amType())

pkt.setDestination(0)

print "Delivering " + str(msg) + " to 0 at " + str(t.time() + 3);

pkt.deliver(0, t.time() + 3)

for i in range(20):

t.runNextEvent()
```

## 3. Langages supportés python et C[++]

L'avantage de python est qu'il est facile à écrire et peut être utilisé de manière interactive. L'inconvénient est qu'il a un coût important. Python/C événements de transition coûtent environ 100 pour cent si cela signifie qu'il exécute une moitié de la vitesse. En outre, il est surtout utile pour passer d'un code avec le débogueur standard.

Tossim supporte le langage C[++] donc. La plupart des scripts des deux (Python et C[++]) reste les même, les deux grandes exceptions peuvent être l'injection de paquets et l'inspection des variables. Habituellement, Python et C[++] donnent le même aspect. [28]

## VI. Performance et analyse de la sécurité

### 1. Analyse et discussion du modèle proposé

#### 1.1 Analyse en termes de performance

Le schéma d'établissement de clé par-paire mise en place a la charge de calcul suivante. Les deux nœuds qui veulent établir la clé par-paire doivent vérifier un code d'authentification de message à partir de tous les voisins et évaluer également une fonction pseudo-aléatoire pour créer la clé par-paire. Le message ACK comporte deux champs, un pour l'ID du nœud et un pour le MAC. Le message HELLO inclut seulement l'ID du nœud. L'espace de stockage requis est seulement pour une clé $K_{in}$. Par conséquent, nous pouvons conclure que la charge de calcul, la communication, la surcharge de stockage pour établir la clé par-paire pour notre schéma est petite.

#### 1.2 Analyse en termes de sécurité

Ce schéma est très efficace en terme de sécurité grâce à la clé par-pairs établie entre deux nœuds, que, même si une clé est compromise, son effet est localisé et l'ensemble du réseau est protégé et loin d'être compromis.

Le régime a été analysé dans la défense contre les divers types d'attaques. Aussi sa survie a été discutée. Quand il y a une attaque et un nœud est compromise, l'adversaire prend avantage de cela et peut lancer des attaques en utilisant les nœuds compromis. Si ce nœud compromis est détecté, le modèle de régénération de clés "re keying" peut efficacement révoquer le nœud attaqué à partir du réseau.

#### 1.3 Connectivité

La connectivité se définie comme la probabilité qu'un nœud puisse établir un lien sécurisé avec l'ensemble de ses voisins. Les protocoles de gestion de clés déterministe comme le cas de notre modèle et LEAP se basent sur la génération de clé initiale et qui achève une connectivité totale (100%) avec un cout faible en matière d'espace de stockage mémoire, contrairement aux protocoles probabilistes qui nécessitent un cout plus élevé en termes de stockage mémoire, et leur connectivité dépond de la taille du trousseau de clés pré chargés dans les nœuds. La probabilité que deux nœuds voisins X et Y partagent un secret varie selon la taille des sous-ensembles de secrets partagés et selon la taille du réseau. La probabilité d'établir un lien sécurisé dépend de la probabilité de partager un secret. Notre modèle comme LEAP est basé sur le pré déploiement d'une même clé initiale à tous les nœuds. Ce mécanisme offre la simplicité d'implémentation et offre une connectivité totale du réseau.

#### 1.4 Complexité en mémoire

Dans notre schéma à l'instar du LEAP [5], un nœud a besoin de garder quatre types de clés. Si un nœud a *D* voisins, il a besoin de stocker une clé individuelle, *D* clés par paire, *D* clés de cluster, et une clé de groupe. Dans un réseau de capteurs du débit de transmission par paquets est généralement faible. Par exemple, les lectures peuvent être générées et transmises périodiquement, et les informations de de routage peuvent être échangées moins souvent. Ainsi, un nœud pourrait stocker un porte-clés d'une longueur raisonnable. Soit *L* le nombre de clés d'un nœud stockés pour sa chaîne. Ainsi, le nombre total de clés qu'un nœud stocke est : $L + D + 2D + 1 + 1$

        *1* ⟵ *Clé privée, 1* ⟵ *Clé de groupe, D* ⟵ *Clés de cluster*
        *2d* ⟵ *Clés par-paire, L* ⟵ *F Fonction*

Bien que l'espace de mémoire soit une ressource très rare pour la génération actuelle de nœuds de capteurs (4 Ko de RAM dans un Berkeley Mica Mote), pour un degré *d* raisonnable, le stockage ne représente pas un problème dans notre système. Par exemple, quand *d = 20* et *L = 20*, un nœud stocke 82 clés (Un total de 656 octets lorsque la taille de clé est 8 octets). Dans l'ensemble, nous concluons que le modèle proposé est évolutive et efficace dans la communication et le stockage.

       *1.5 La résistance et défense contre les diverses attaques*

Le protocole LEAP suppose que les phases d'initialisation et d'installation de clés s'exécutent en un temps maximal Tmin et qu'un attaquant prendra au minimum un temps T > Tmin pour compromettre ou capturer un nœud et récupérer les informations contenues dans sa mémoire. En plus de mécanisme proposé par Zhou et al [5] basé sur la révocation du nœud et la suppression de la clé Kinit de la mémoire de tous les nœuds et la régénération de clés symétriques entre chaque paire de nœuds, notre modèle renforce la sécurité par la mise en place de deux approches pour la détection du nœud compromis discutées précédemment. Le contrôle des informations de routage est authentifié par le schéma local de diffusion d'authentification qui empêche la plupart des attaques externes. Les éventuelles attaques qu'un adversaire peut lancer dans l'espoir de créer des boucles de routage, récupérer le trafic réseau ou générer des messages d'erreur sont :

- Spoof
- Alter
- Replay routing information
- Attaque Selective forwarding
- Attaque HELLO Flood
- Attaque de clonage de nœud

Ce schéma ne peut pas empêcher l'adversaire de lancer ces attaques mais il peut contrer ou minimiser leurs conséquences. Cet effet de localisation aide aussi à détecter ces types d'attaques.

L'attaque Alter : Elle peut également être détectée comme le nœud d'envoi peut entendre le message étant modifiée au cours de sa phase de transmission par le nœud compromis. Il est également intéressant de mentionner que si un nœud est compromise et détecter, le système de régénération de clés peut efficacement révoquer le nœud du réseau (Voir section 4).

L'attaque Hello Flood : Un adversaire peut essayer d'envoyer un message Hello à chaque nœud avec une puissance d'émission élevée pour convaincre tous les nœuds voisins. Ici l'attaque Hello flood ne sera pas réussi au-delà de la phase de découverte de voisinage parce que chaque nœud accepte les paquets seulement des voisins qui sont authentifiés. La même chose pour l'attaque clonage de nœud ne peut pas aller aussi au-delà de la phase de découverte de voisinage.

Sybil attaque : Dans l'algorithme, un MAC de l'identifiant du nœud, son niveau, et l'identifiant de son père est calculé pour authentifier l'expéditeur et le destinataire. Par conséquent, un nœud ne peut pas jouer un rôle d'autres nœuds.

L'attaque Node capture: Quand un nœud est capturé, cela ne porte pas atteinte à ses voisins. En effet, après un nœud est capturé, ce que peut faire un attaquant? Depuis qu'il a la clé partagée avec la station de base, il peut envoyer de fausses informations de la station de base. Ce dernier peut disposer d'un mécanisme pour vérifier le comportement de nœuds émetteurs. L'attaquant a également accès aux clés des fils de la victime qui lui permet par la suite d'envoyer des messages inutiles à ces fils pour consommer leur énergie et leur causer épuisement de la batterie.

Notre modèle ainsi que LEAP empêche un adversaire de lancer l'attaque trou de ver « Wormhole » ou l'attaque « Sinkhole » parce que le nœud connaît tous ses voisins après l'étape de découverte de voisin. Depuis un attaquant ne peut pas dériver des clés à partir des initiales clés initiales ou des clés maîtresses. Résilience contre la capture de nœud : ou résistance contre la capture de nœud, cette métrique mesure comment le RCSF est compromis quand un nœud est compromis, et l'influence de ce nœud sur la sécurité du réseau. Dans notre schéma contrairement au protocole LEAP qui suppose que dans Tmin le nœud ne peut pas être compromis, on a mis en place deux mécanismes pour contrer l'effet du compromis du nœud. D'autres travaux cités dans la section 4 [22] [23] [24] [25] [26] ont traités et proposer des modèles pour détecter le nœud compromis.

## 2. Expérimentation

### 2.1 Analyse de temps de génération de la clé par-paire

Comme le temps disponible pour la génération de clé par-paire est court, on a analysé le temps pour la génération de la clé par-paire avec différentes nombre de nœuds. L'intérêt principal était de vérifier si la densité du réseau a un effet sur la génération réussite de la clé par-paire dans le temps disponible. Nous avons aperçu que l'algorithme génère avec succès de clés de par-paire avec différents nœuds dans le temps disponible. Il a été remarqué également que chaque paire de voisin génère avec succès la même clé par-paire qui est identifié de manière unique par la paire de nœud. L'évaluation a été réalisée avec une différence de deux, cinq et dix nœuds et répéter dix fois pour chacune. Presque le même résultat a été remarqué pour chaque modèle simulé avec une génération réussite de la clé par-paire.

#### 2.1.1 Premier Modèle expérimenté

- Exécuté avec une différence de deux nœuds répétée dix fois

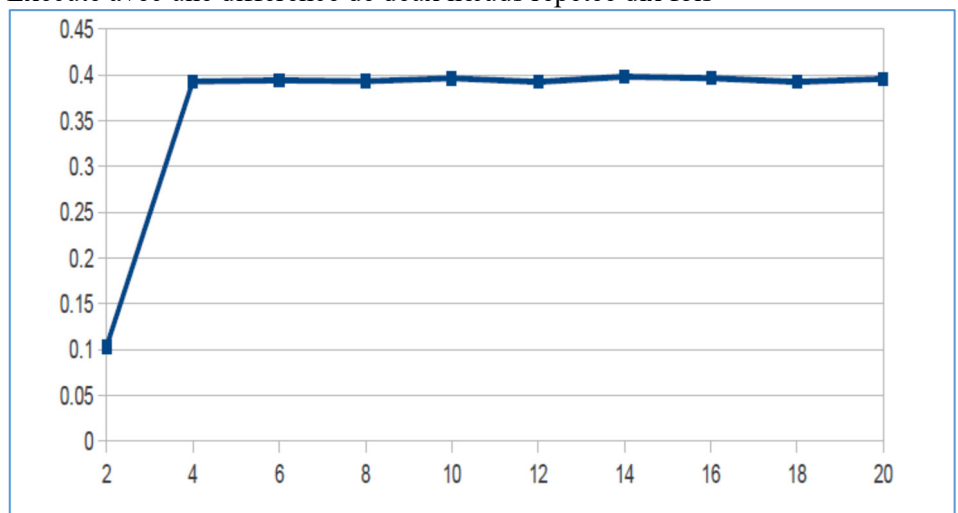

Figure 8. Analyse du temps de Génération de clé par-paire

#### 2.1.2 Second Modèle expérimenté

- Exécuté avec une différence de cinq nœuds répétée dix fois

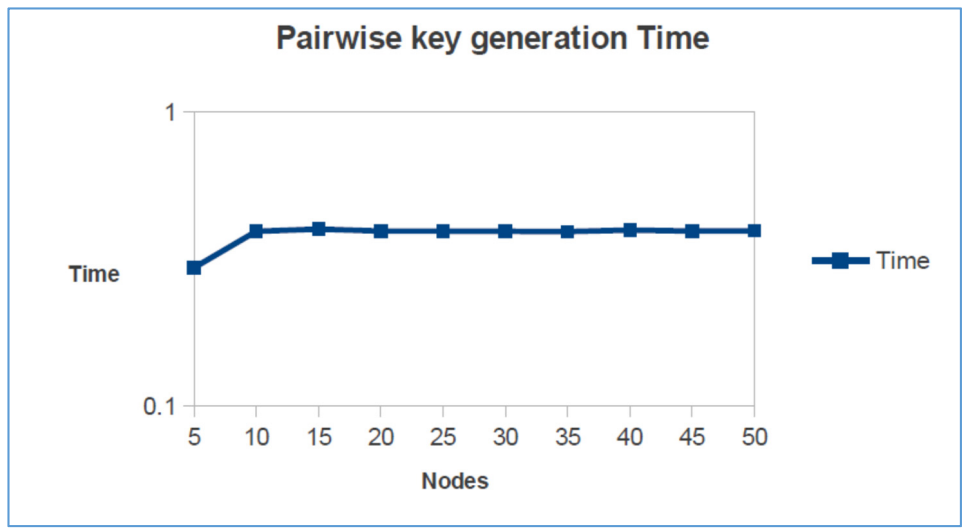

Figure 9. Analyse du temps de Génération de clé par-paire

*2.1.3 Troisième modèle expérimenté*

- *Exécuté avec une différence de 10 nœuds dix fois de suite*

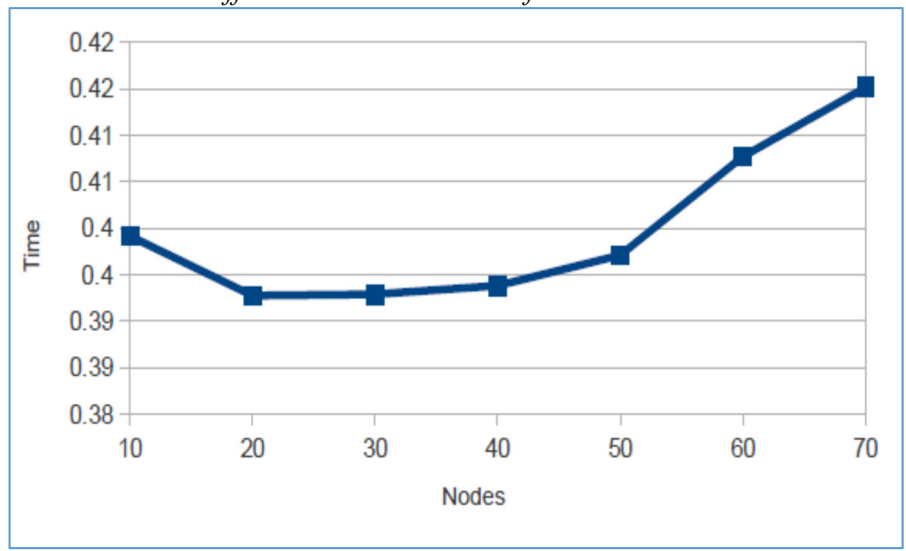

Figure 10. Analyse du temps de Génération de clé par-paire

## 2.2 Analyse du temps de génération de la clé individuelle

L'étape la plus importante était de vérifier la réussite de la génération de clé par-paire. Mis à part, le temps de génération de clés individuelles pour les différents modèles de nœuds a également été analysé. Il a été constaté que chaque paire de nœud succès générer une clé individuelle unique.

### 2.2.1    Premier modèle d'expérimentation

- *Exécuté avec un échantillon de dix nœuds*

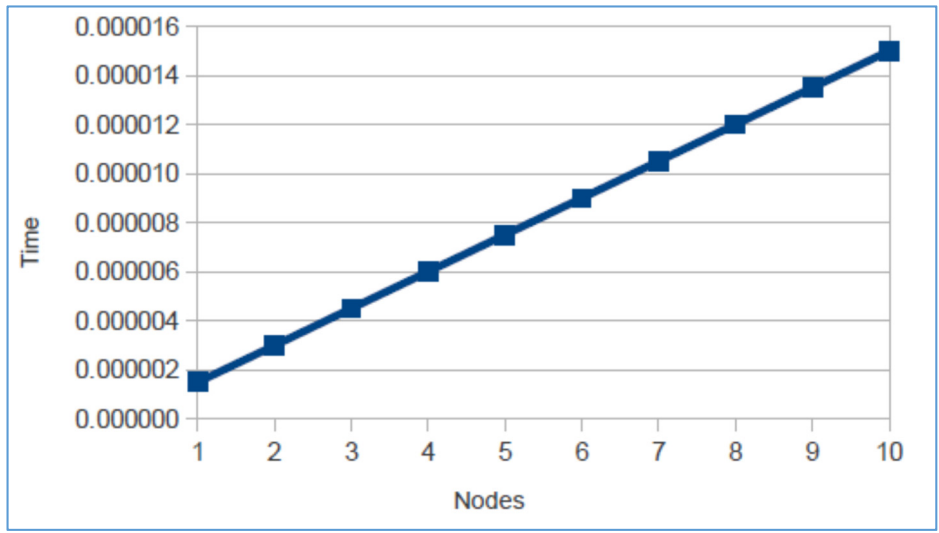

Figure 11. Analyse du temps de Génération de clé individuelle

*2.2.2 Second model d'expérimentation*

- Exécuté avec un échantillon de 20 nœuds

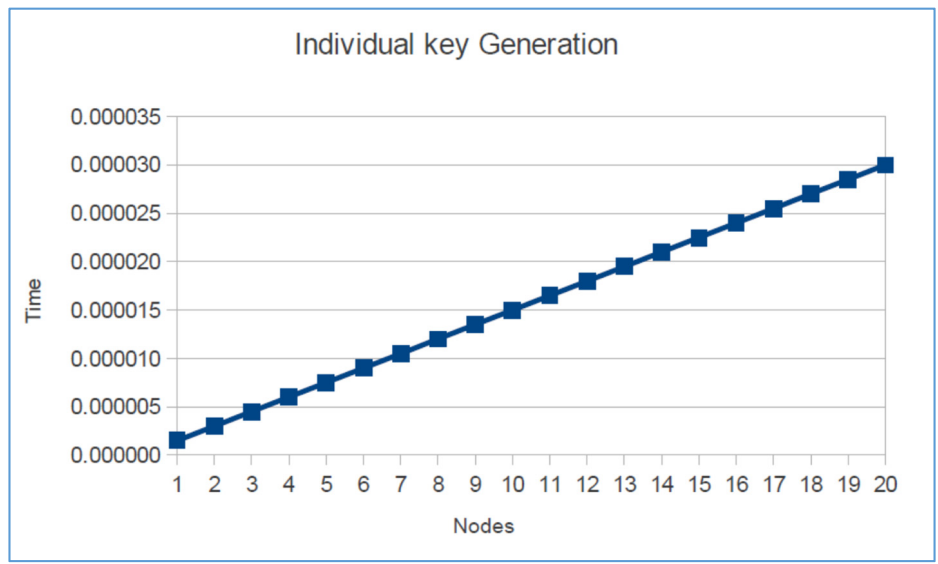

Figure 12. Analyse du temps de Génération de clé individuelle

## 2.3 Evolutivité

Pour évaluer le passage à l'échelle (évolutivité) de notre proposition, nous avons réalisé différentes expérimentations sur des RCSFs en variant le nombre de nœud de 10 à 100 nœuds, nous avons calculé le nombre maximum de messages pouvant être reçus par un nœud en fonction du nombre des nœuds capteurs dans le réseau. La figure 13 résume ces résultats.

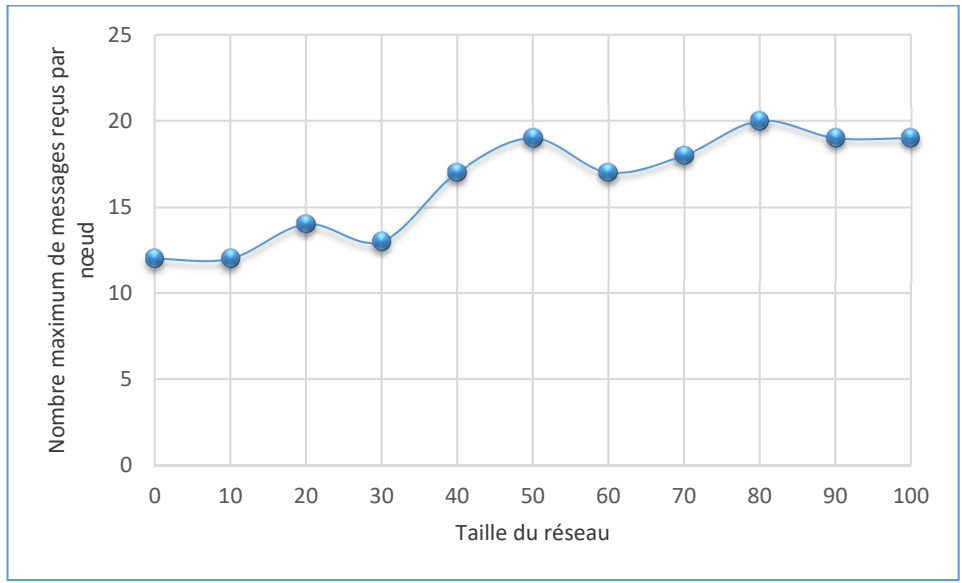

Figure 13. Nombre maximum de message reçu par un nœud Vs la taille du réseau

Nous pouvons observer que notre modèle n'engendre pas des coûts supplémentaires lorsque le nombre de nœuds dans le réseau augmente. Le nombre de message reçu par les nœuds reste stable par rapport à la densité du réseau.

### 2.4 Consommation moyenne d'énergie

En utilisant l'outil PowerTossim, nous avons évalué la consommation moyenne d'énergie d'un nœud capteur durant les phases de découverte de voisinage et d'installation des clés de session. Cette énergie est calculée sur la base des instructions exécutées pour les opérations cryptographiques (calcul du MAC par l'émetteur, vérification du MAC par le récepteur, calcul de la clé individuelle, calcule de la clé par-paire) et pour les opérations radio (émission et réception des messages de découverte de voisinage, d'échange des messages Hello et MAC). La figure 13 illustre la variation de l'énergie consommée en fonction du nombre de nœuds du réseau en présence de nœuds attaquants.

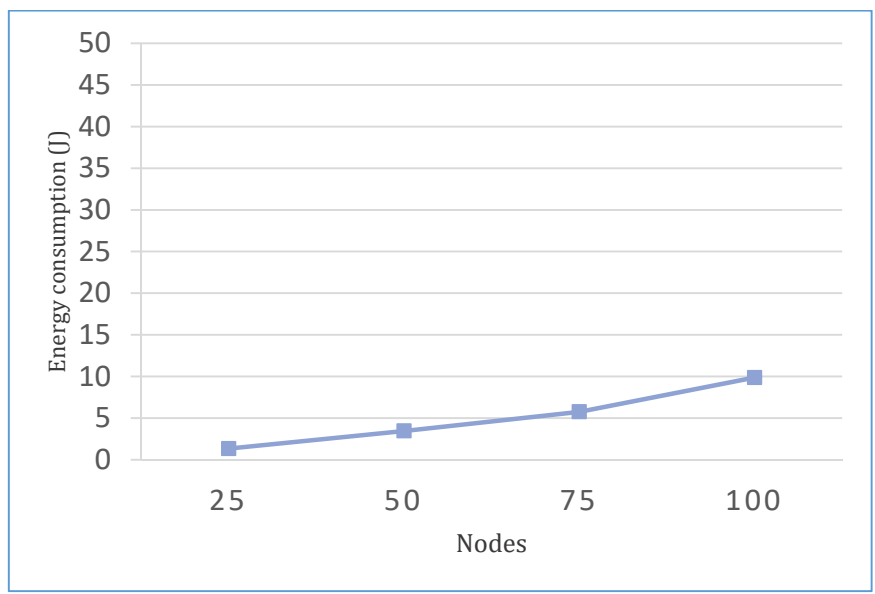

Figure 13. L'énergie consommée

Dans la figure 13, on remarque que la quantité d'énergie dépensée augmente avec la croissance des nœuds et la taille du réseau. Cette augmentation de consommation d'énergie est due principalement au nombre croissant de paquets échangés pour l'établissement, le calcul et la vérification des clés de cryptage par les nœuds capteurs.

## 3. Comparaison et discussion

### *3.1 Méthodes d'analyse et de comparaison*

Dans cette section, nous évaluons la performance de notre solution avec quelques schémas existants dans la littérature. Plusieurs critères sont pris en charge afin de comparer les différentes méthodes de gestion de clés. Nous présentons dans la figure 14 les critères et mesures pour atteindre cette évaluation.

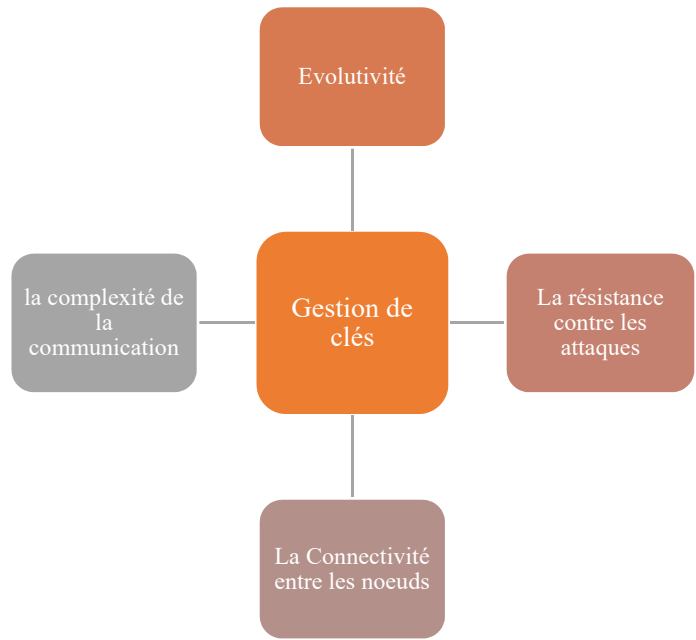

Figure 14. Les critères d'évaluation des méthodes de gestion de clés dans les RCSF

- La complexité de la communication : La méthode de gestion de clés proposée doit prendre en considération le fait que les nœuds ont été déployés pour collecter les informations. Ils ont besoin de leur espace mémoire pour stocker leurs données et de leur énergie embarquée pour assurer leur rôle applicatif.
    1. Faible : le protocole ne nécessite pas beaucoup de ressource en matière de stockage mémoire et consommation d'énergie,
    2. Moyenne : Le protocole nécessite une consommation moyenne en termes d'énergie et mémoire,
    3. Elevé : Le protocole est gourmand en matière de consommation d'énergie et stockage de mémoire.
- la connectivité : la probabilité que deux nœuds (ou plus) partagent une clé.
- La Résilience contre la capture de nœud: cette métrique mesure l'impact d'un nœud compromis sur la sécurité du reste du réseau.
  Nous quantifions cette métrique avec les trois valeurs suivantes :
    1. Bonne résistance : le nœud compromis ne touche que ses voisins (influence locale),
    2. Faible résistance : le nœud compromis affecte ses voisins et aussi quelques non nœuds voisins,
    3. Très faible résilience : si le nœud de l'un compromis conduit à compromettre l'ensemble du réseau.
- L'évolutivité : cette métrique mesure de flexibilité du protocole avec la taille du réseau. En d'autres termes, la métrique montre comment le coût du protocole. L'évolutivité est une mesure très important à considérer lors la conception des algorithmes proposés, en particulier pour les réseaux de capteurs. Pour quantifier l'évolutivité, nous utilisons les valeurs suivantes :
    1. Bonne : le protocole ne induit pas des coûts supplémentaires lorsque le nombre de nœuds dans le réseau augmente,

2. Juste : le protocole induit des coûts raisonnables lorsque le nombre de nœuds augmente,
3. Limité : le coût du protocole en fonction du nombre de nœuds.

### 3.2 Comparaison avec d'autres schémas

Dans le Tableau 1, nous avons comparé notre modèle avec des méthodes existantes dans la littérature en se basant sur les critères déjà discutées dans la figure 14 et dans notre papier [27] et les travaux des auteurs [33][34]. Nous remarquons que notre schéma présente la résistance la plus efficace face aux effets produits par les captures de nœud. Notre modèle est équivalent à *LEAP* en termes de passage à l'échelle « L'évolutivité ». Cette équivalence vient du fait que ces deux schémas utilisent des clés symétriques permettant de passer à l'échelle plus facilement. Nous observons aussi qu'en termes de connectivité, notre schéma est équivalent à plusieurs systèmes symétriques, cela à cause de l'utilisation de la phase *NeighbDisc* après le déploiement. N'importe quel nœud peut établir un lien avec son voisin en demandant sa clé publique à la station de base. Notre solution a une communication à faible complexité par rapport aux autres solutions proposées, avec une connectivité de clés équivalente à un (pas de notion de probabilité). Ayant un moyen de complexité de communication faible veut dire que la solution proposée est légère.

| Schémas | Critères de comparaison | | | | |
|---|---|---|---|---|---|
| | Évolutivité | Connectivité | La Résilience contre la capture de nœud: | La complexité de la communication | |
| | | | | Complexité de mémoire | Consommation d'énergie |
| Notre schéma | Bonne | 100% | Bonne | Faible | Faible |
| TinyPK [19] | Juste | La probabilité que deux nœuds partage une clé $P_1$ | Très faible | Moyenne | Elevé |
| SPINS [15] | Limité | 100% | Faible | Faible | Moyenne |
| TinySec [16] | Juste | La probabilité que deux nœuds partage une clé $P_2$ w $P_1$ | Faible | | Elevé |
| LEAP [5] | Bonne | 100% | Bonne | Faible | Faible |
| PKKE [20] | Juste | La probabilité que deux nœuds partage une clé $P_1$ | Faible | Elevé | Elevé |

Tableau 1. Comparaison avec d'autres schémas de sécurité dans les RCSF

La comparaison élaborée dans cette partie conclue que notre méthode a un niveau de sécurité élevé contre les attaques par rapport à ceux décrits dans notre revue de la littérature. Les résultats ont montré que notre modèle à un coût de stockage des clés faibles, est connectivité de 100%. Notre cryptage des données est symétrique, le temps consommé dans le coût et l'énergie est équivalente à d'autres méthodes symétriques comme SPINS et LEAP. La solution que nous avons proposée présente une connectivité totale, une génération réduite du nombre et de la taille des paquets échangés.

## VII. Conclusion

La sécurité est une des enjeux majeure dans les réseaux de capteurs sans fil en raison de leurs domaines d'applications. Ils peuvent être implémentés dans un des systèmes très critiques comme les hôpitaux, les aéroports, les applications militaires, les alarmes antivol, le contrôle de l'environnement, les maisons intelligentes et la surveillance du trafic. L'objectif principal de ce travail est d'aborder la question de la sécurité dans les réseaux de capteurs. Les réseaux de capteurs sont vulnérables contre les attaques externes et internes en raison de leurs caractéristiques uniques, ils sont limités en terme de capacité de calcul, de communication et de mémoire. Les mécanismes de sécurité classique ne sont pas adaptés pour les réseaux de capteurs en raison de contraintes discutés avant. Ainsi, les études des systèmes de gestion de clés dans les RCSF nous ont mené à conclure que l'utilisation des systèmes de clés symétriques partagées sont les plus appropriés pour ce type de réseau. Une observation importante pour tout type de système de gestion des clés est qu'un mécanisme unique de chiffrement ne convient pas pour satisfaire les exigences de sécurité différentes. Dans ce travail différent algorithmes de cryptage pour les RCSF ont été étudié d'une manière approfondie, afin de proposer un système de gestion des clés basé sur le partage des clés symétriques. Les résultats prouvent que notre modèle montre de bonnes performances en termes de connectivité, gestion de ressources mémoire et résistance contre les attaques. La croissance rapide des réseaux de capteurs dans différents domaines et spécialement ceux critiques a attiré l'attention d'un grand nombre de chercheurs pour travailler sur la partie sécurité. Dans les futurs travaux nous essayerons d'adopter notre algorithme pour l'adapter pour les réseaux 6lowPan.

## Les références

Nom du document : Chapitre-Etude et développement d'un protocole pour sécuriser les communications des RCSF (1)
Répertoire : C:\Users\maleh\Documents
Modèle : C:\Users\maleh\AppData\Roaming\Microsoft\Templates\Normal.dotm
Titre :
Sujet :
Auteur : Yassine MALEH
Mots clés :
Commentaires :
Date de création : 26/12/2015 17:14:00
N° de révision : 45
Dernier enregistr. le : 29/12/2015 14:30:00
Dernier enregistrement par : MALEH YASSINE
Temps total d'édition : 915 Minutes
Dernière impression sur : 30/12/2016 19:37:00
Tel qu'à la dernière impression
    Nombre de pages : 28
    Nombre de mots : 12 560 (approx.)
    Nombre de caractères : 69 083 (approx.)